\documentclass[manuscript]{aastex}

\slugcomment{To appear in an upcoming edition of ApJ}

\shorttitle{Intermediate Mass Black Hole in NGC 404}
\shortauthors{Nyland et al.}

\begin{document}

\title{The Intermediate Mass Black Hole Candidate in the Center of NGC~404: New Evidence from Radio Continuum Observations}


\author{Kristina Nyland\altaffilmark{1,2},\email{knyland@nmt.edu} Josh Marvil\altaffilmark{1,2}, J. M. Wrobel\altaffilmark{2}, Lisa M. Young\altaffilmark{1}, B. Ashley Zauderer\altaffilmark{3}}

\altaffiltext{1}{New Mexico Tech, Department of Physics, 801 Leroy Place, Socorro, NM 87801}
\altaffiltext{2}{National Radio Astronomy Observatory, Socorro, NM 87801, USA}
\altaffiltext{3}{Harvard University, 60 Garden Street, Cambridge, MA 02138, USA}

\begin{abstract}
We present the results of deep, high-resolution, 5~GHz Expanded Very Large Array (EVLA) observations of the nearby, dwarf lenticular galaxy and intermediate mass black hole candidate ($M_{\mathrm{BH}}$~$\sim$~4.5~$\times$~10$^5$~M$_{\sun}$), NGC~404.  For the first time, radio emission at frequencies above 1.4~GHz has been detected in this galaxy.  We found a modestly resolved source in the NGC~404 nucleus with a total radio luminosity of 7.6~$\pm$~0.7~$\times$~10$^{17}$~W~Hz$^{-1}$ at 5~GHz and a spectral index from 5 to 7.45~GHz of $\alpha$~=~$-$0.88~$\pm$~0.30.  NGC~404 is only the third central intermediate mass black hole candidate detected in the radio regime with subarcsecond resolution.  The position of the radio source is consistent with the optical center of the galaxy and the location of a known, hard X-ray point source ($L_{\mathrm{X}}$~$\sim$~1.2~$\times$~10$^{37}$~erg~s$^{-1}$).  The faint radio and X-ray emission could conceivably be produced by an X-ray binary, star formation, a supernova remnant or a low-luminosity AGN powered by an intermediate mass black hole.  In light of our new EVLA observations, we find that the most likely scenario is an accreting intermediate mass black hole, with other explanations incompatible with the observed X-ray and/or radio luminosities or statistically unlikely.


\end{abstract}


\section{Introduction}
\label{sec:intro}
It is widely accepted that supermassive black holes (SMBHs) with masses in the range 10$^{6}$~$-$~10$^{9}$ M$_{\sun}$ are commonplace in galactic nuclei \citep{kor_rich95}.  However, knowledge of the demographics and characteristics of black holes (BHs) in the centers of less massive galaxies is currently lacking.  It is possible that many lower-mass galaxies simply host scaled-down versions of the central SMBHs in more massive galaxies, with masses $<$~10$^{6}$~M$_{\sun}$ \citep{mckernan11}.  Such low BH masses would occupy the {\it intermediate mass black hole} (IMBH) regime \citep{vandermarel04}, which comprises the population of BHs with masses between those of stellar-mass BHs and SMBHs (10$^{2}$~$-$~10$^{6}$~M$_{\sun}$).  The detection and study of IMBHs promises to address a number of key astrophysical issues.  For example, detections of IMBHs in low-mass galaxies would provide an excellent diagnostic tool for probing the formation mechanisms of `seed' BHs \citep{volonteri10, vanwass10} as well as more exotic phenomena, such as gravity waves \citep{hughes09}.  Although there are a number of promising IMBH candidates in the centers of external galaxies, these detections have remained scarce, challenging and controversial \citep{greene07b, desroches09, barth08}.  As a complication, it has been suggested that galaxies with total masses $<$ 10$^{10}$ M$_{\sun}$ may be dominated by massive nuclear star clusters (NSCs) rather than scaled-down versions of SMBHs \citep{ferrarese06}.  There does exist a significant population of galaxies which appear to contain {\it both} a central massive BH and an NSC \citep{graham09, seth08}.  The fraction of galaxies harboring both an NSC and a central IMBH candidate is currently unknown, but a number of studies in recent years have explored such objects.  Examples include some globular clusters which may actually be the stripped-down nuclei of former dwarf galaxies, such as $\omega$~Cen in the Galaxy \citep{noyola10, vandermarel10}, G1 in M31 \citep{gebhardt02, gebhardt05, ulvestad07, kong10}, M54 in the Sagittarius dwarf galaxy \citep{ibata09, wrobel11} and ESO~243-49~HLX-1 \citep{farrell12}, though these cases have been controversial.  Other well-known examples include the galaxies NGC~4395 \citep{filippenko03, peterson05}, Pox~52 \citep{barth04, thornton08}, Henize~2-10\footnote{The relatively uncertain central BH mass estimate in this dwarf galaxy of a few $\times$ 10$^6$ M$_{\sun}$ is not technically in the IMBH regime, but the error bars on this mass estimate span over an order of magnitude.  Thus, more robust $M_{\mathrm{BH}}$ measurements may ultimately establish Henize 2-10 as an IMBH candidate.} \citep{reines11}, NGC~1042 \citep{shields08}, NGC~3621 \citep{barth09}, NGC~4178 \citep{satyapal09}, NGC~3367 and NGC~4536 \citep{mcalpine11} and NGC~404 \citep{seth10, binder11}.   

NGC~404 is a nearby (D~=~3.1~Mpc), dwarf S0 galaxy with a complex nuclear environment.  Unlike typical lenticular galaxies, which tend to be gas depleted with little signs of star formation (SF) and populate the red sequence on the color magnitude diagram \citep{baldry04}, NGC~404 appears to have been rejuvenated with cold gas during a merger $\sim$1~Gyr ago \citep{delrio04, bouchard10} and is currently undergoing low-level SF in a faint outer disk \citep{thilker10} at a radius of a few hundred arcseconds (a few kpc).  SF may have also occurred recently in the NGC~404 nucleus, as indicated by the detection of an NSC with an effective radius of 0.7$\arcsec$ (10~pc) and a mass of 1.1~$\pm$~0.2~$\times$~10$^7$~M$_{\sun}$ \citep{ravindranath01, seth10}.  Within the confines of the NSC, NGC~404 may also harbor a central IMBH.  Compelling evidence in support of such an IMBH includes high resolution HST and ground-based adaptive optics dynamical studies \citep{seth10}.  Seth et al.\ (2010) reported dynamical BH mass estimates of $M_{\mathrm{BH}}$~$<$~10$^{5}$~M$_{\sun}$ (stellar dynamical upper limit) and $M_{\mathrm{BH}}$~=~4.5$^{+3.5}_{-2.0}$~$\times$~10$^{5}$~M$_{\sun}$ (gas dynamical estimate).

Observations of the NGC~404 nucleus at a variety of wavelengths have provided substantial evidence for the presence of low-level accretion onto the candidate IMBH.  Hydrogen recombination line ratios have revealed that NGC~404 harbors a type~2 LINER\footnote{A LINER is defined as a ``low-ionization nuclear emission-line region;" type 2 LINERs lack the characteristic broad permitted lines present in type 1 AGNs.  See Ho 2008 for review.} \citep{ho97}.  Although the presence of a low-luminosity AGN (LLAGN) in all type~2 LINERs is a controversial subject, many type~2 LINERs are known to harbor accreting, massive BHs \citep{ho2008}.  The {\it Chandra} analysis of NGC~404 by Binder et al.\ (2011) revealed a central, variable, hard (2$-$10 keV) X-ray point source with a luminosity of $1.2^{+0.7}_{-0.4}$~$\times$~10$^{37}$~erg~s$^{-1}$ and a power-law spectrum ($\Gamma$~=~1.88).  Although the weak, hard X-ray emission could be generated by a single X-ray binary (XRB) or an LLAGN, Binder et al.\ (2011) argued that the X-ray spectral shape and luminosity were most consistent with other observed LLAGNs and inconsistent with XRBs.  Data at other wavelengths has also provided evidence for the presence of an LLAGN.  In 2004, Satyapal et al.\ reported that the nuclear mid-IR emission lines of NGC~404 were consistent with bona fide AGNs in their sample.  Near-IR data from Gemini have revealed unresolved and potentially variable hot dust emission, which is characteristic of Seyfert nuclei \citep{seth10}.  In the UV regime, clear, yet shallow absorption features have indicated the presence of O stars in the center of NGC~404, along with an additional continuum component attributable to either an LLAGN or a population of less massive stars \citep{maoz98}.  UV variability observed over a 9 year period also supports the existence of an LLAGN \citep{maoz05}.

A more thorough analysis of the nuclear engine of NGC~404 may ultimately help improve our understanding of SMBH growth and the relationship between massive BH evolution and the evolution of the host galaxies in which they reside.  High-resolution, deep, interferometric radio observations offer the ability to easily detect and pinpoint emission from weakly accreting, low-mass BHs \citep{maccarone04}.  Yet, besides NGC~404, only two other candidate IMBHs, NGC~4395 \citep{hoandulvestad01, wrobel01, wrobelandho06} and GH~10 \citep{greene06, wrobel08}, have subarcsecond-scale radio continuum detections.  In this paper, we present an analysis of our recent Expanded Very Large Array (EVLA)  detection of parsec-scale radio emission in the NGC~404 nucleus and the implications it has on the status of NGC~404 as an IMBH candidate.  In Section~\ref{sec:obs} we describe our observations and data reduction procedure.  We discuss the possible physical interpretations of our EVLA observations in Section~\ref{sec:results} and summarize our work in Section~\ref{sec:summary}.

\section{Observations}
\label{sec:obs}
We observed NGC~404 with the NRAO\footnote{The National Radio Astronomy Observatory is a facility of the National Science Foundation operated under cooperative agreement by Associated Universities, Inc.} EVLA \citep{perley11} at $L$-band (1-2~GHz) and $C$-band (4-8~GHz).  The $L$-band observations were carried out in the B-configuration on April 11 and 30, 2011 (project ID: 10C-145) over 2 hours with a single 1~GHz-wide baseband centered at 1.5~GHz.  The WIDAR correlator was configured with 8 spectral windows, each of which contained 64 channels with a frequency resolution of 2~MHz.  These observations were phase-referenced to the calibrator J0119+3210 every 6 minutes with a switching angle of 4$\degr$.  The positional accuracy of our phase calibrator was $<$0.002$\arcsec$.  The calibrator 3C48 was used to set the amplitude scale to an accuracy of 3\% and calibrate the bandpass.  Four out of the 27 antennas experienced severe hardware malfunctions and were removed from the dataset.  $L$-band data from the remaining antennas were flagged and calibrated with the Astronomical Image Processing System (AIPS) using standard procedures.  We manually inspected each baseline and flagged the data using the AIPS task SPFLG to excise the heavy radio frequency interference (RFI) present in the $L$-band frequency range. 

The $C$-band observations were carried out in the A-configuration on July 9, 2011 (project ID: TDEM0009) over 30 minutes with two 1~GHz-wide basebands centered at 5 and 7.45~GHz.  The WIDAR correlator was configured with 16 spectral windows split across the two basebands, each of which contained 64 channels with a frequency resolution of 2~MHz.  These observations were phase-referenced to the calibrator J0111+3906 every 5 minutes with a switching angle of 3$\degr$.  The positional accuracy of our phase calibrator was $<$0.002$\arcsec$.  The calibrator 3C48 was used to set the amplitude scale to an accuracy of 3\% and calibrate the bandpass.  We flagged, calibrated and imaged the $C$-band data with the Common Astronomy Software Applications package (CASA).  The WIDAR correlator malfunctioned during our observations and caused large amplitude jumps at scan boundaries, affecting intermediate frequencies B and C.  We were able to correct this problem by applying switched power during data import to CASA and re-calibrating the gain.  RFI was minimal in our chosen basebands and only a few isolated frequencies which displayed bright, narrowband RFI required flagging.  Three out of the twenty-seven antennas contained significantly corrupted data and were partially removed when necessary. 

\begin{figure}
\begin{center}
\includegraphics[scale=.4]{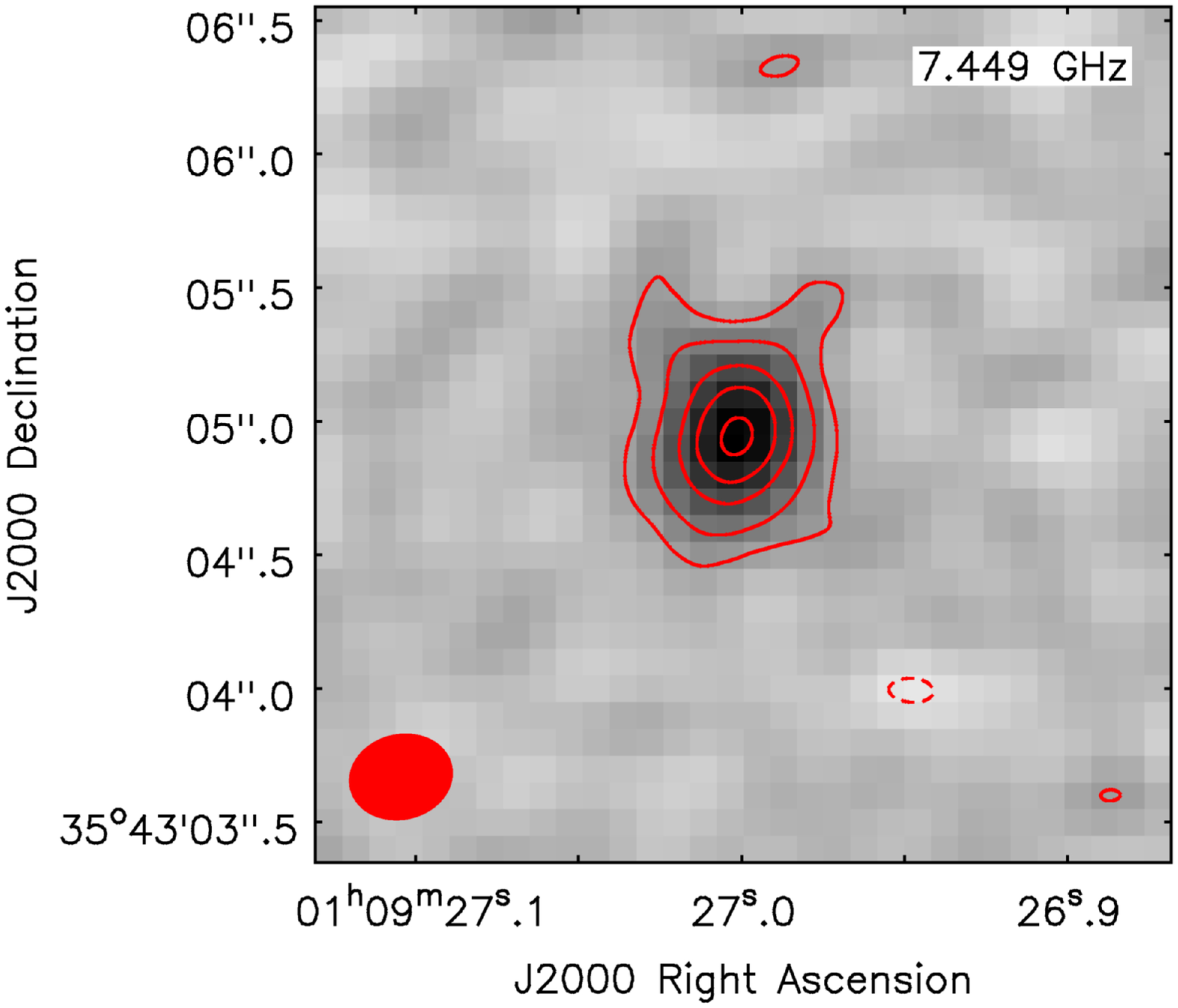}
\includegraphics[scale=.4]{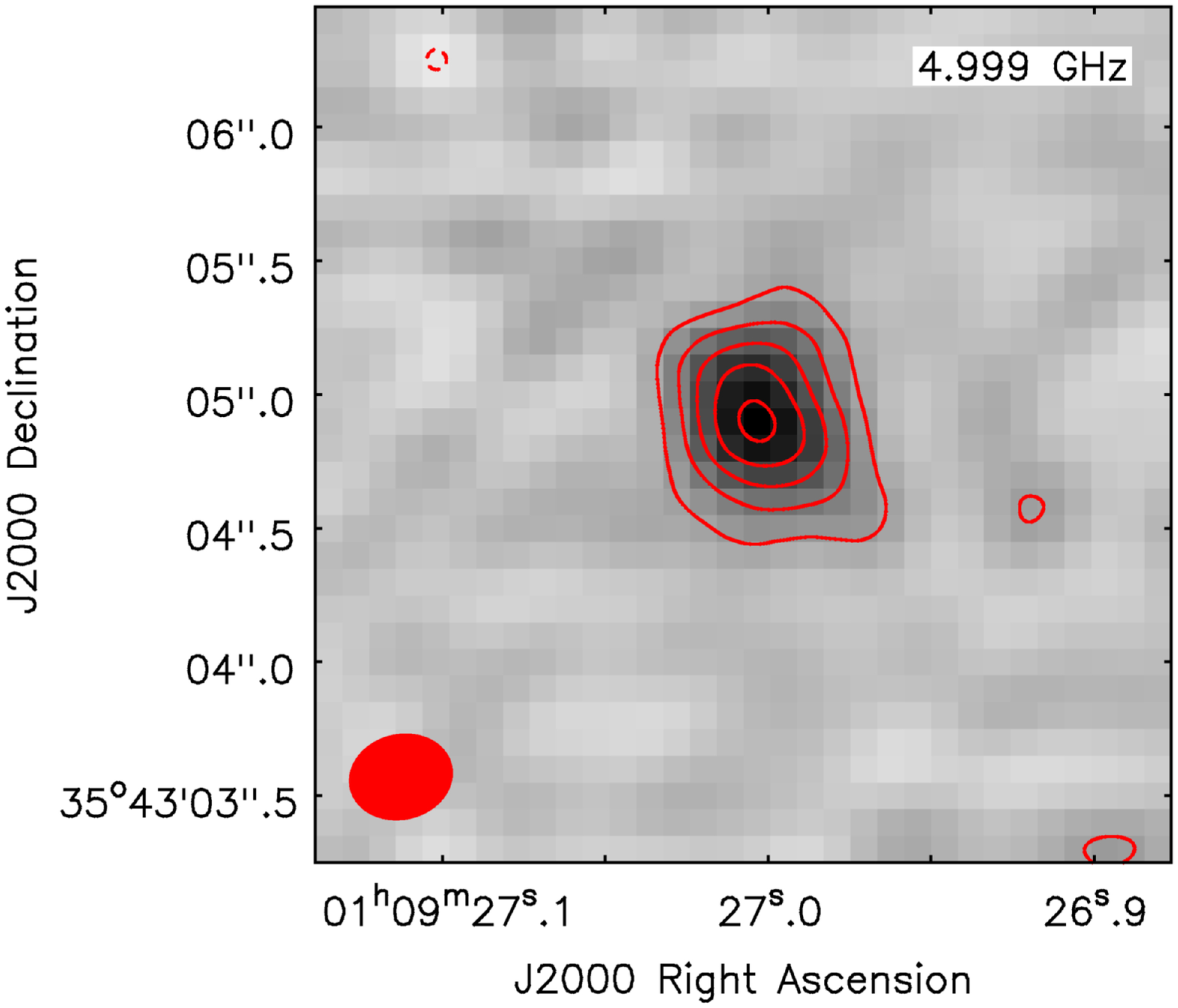}
\includegraphics[scale=.4]{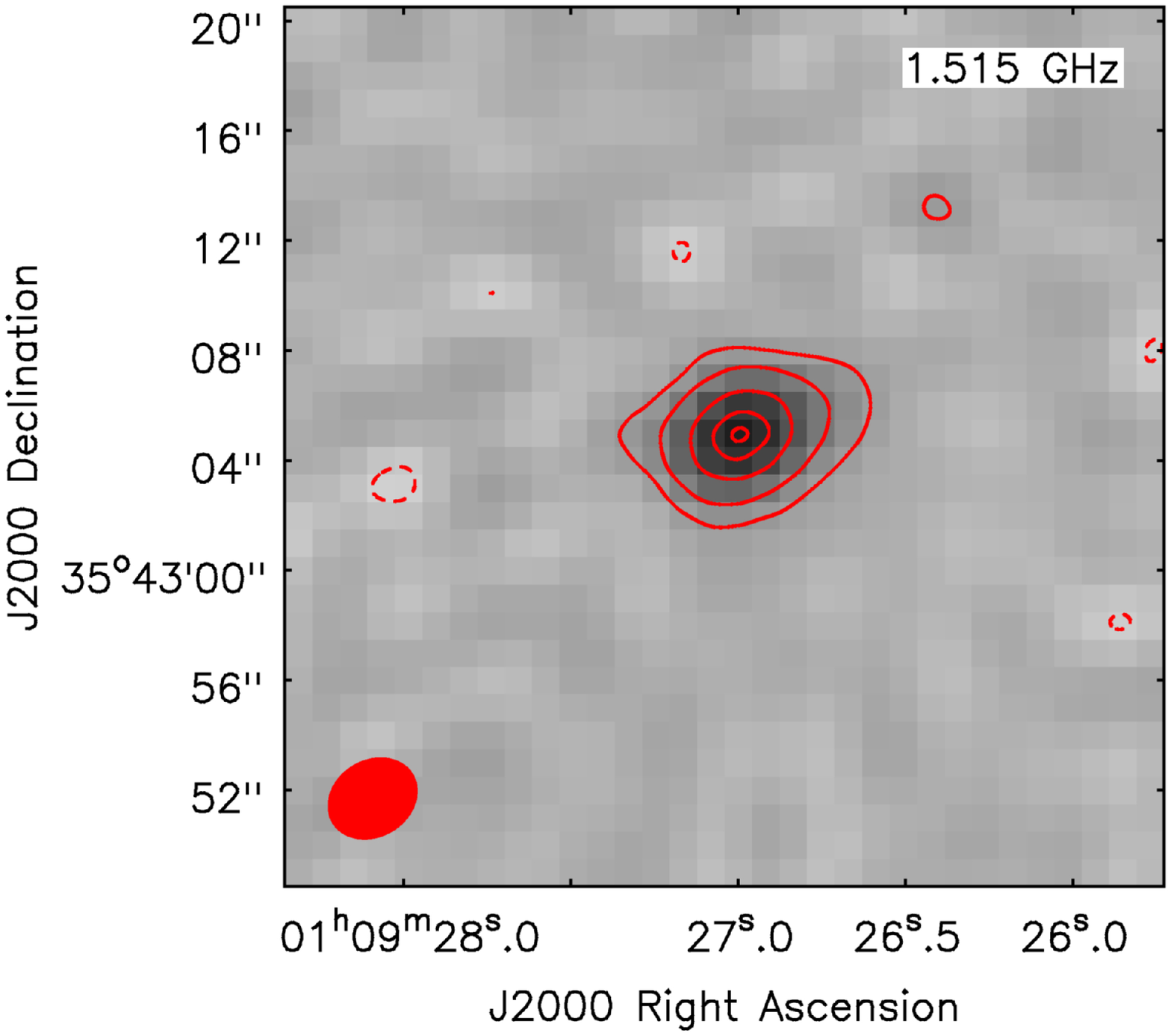}
\caption{NGC~404 radio continuum images with contours at 7.45~GHz, 5~GHz and 1.5~GHz.  The EVLA beam is the filled circle shown on the lower left in each image.  For the 7.45~GHz image, the relative contour levels are [-3,~3,~6,~10,~13,~16] and the unit contour level is 11~$\mu$Jy~beam$^{-1}$.  For the 5~GHz image, the relative contour levels are [-3,~3,~6,~9,~12,~15] and the unit contour level is 16~$\mu$Jy~beam$^{-1}$.  For the 1.5~GHz image, the relative contour levels are [-3,~3,~8,~20,~32,~39] and the unit contour level is 55~$\mu$Jy~beam$^{-1}$.  Note that 1$\arcsec$ = 15~pc.  \label{fig:radio}}
\end{center}
\end{figure}

\begin{figure}
\epsscale{0.8}
\plotone{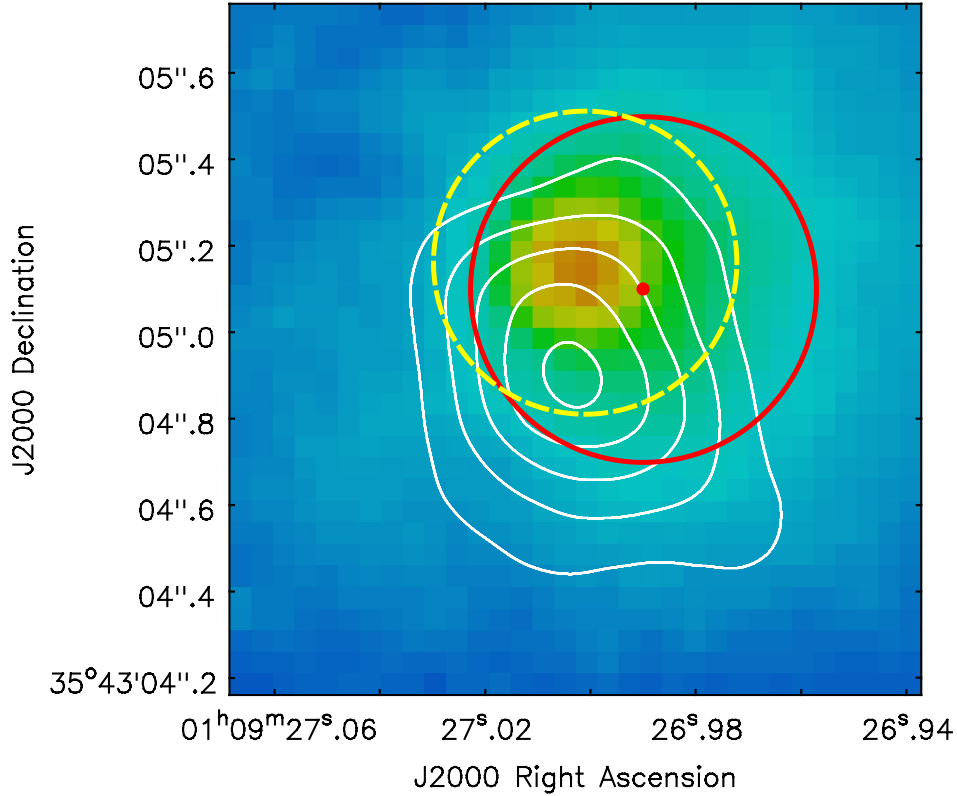}
\caption{$HST$ WFPC2 F814W image of the NGC~404 nucleus (Hubble Legacy Archive) with the EVLA 5~GHz contours overlaid in white.  The relative contour levels are [-3,~3,~6,~9,~12,~15] and the unit contour level is 16~$\mu$Jy~beam$^{-1}$.  The red dot denotes the position of the hard X-ray point source \citep{binder11} and the red circle illustrates the $Chandra$ absolute positional uncertainty of 0.4$\arcsec$ (6.0~pc).  The dashed yellow circle shows the $HST$ absolute positional uncertainty of 0.35$\arcsec$ (5.3~pc) after registration with 2MASS point sources.  The 5~GHz source has a positional uncertainty of 0.1$\arcsec$ (1.5~pc).  Within the positional uncertainties, the 5~GHz emission is consistent with the location of the hard X-ray point source and the optical center of NGC~404. \label{fig:overlay}}
\end{figure}

We imaged the $L$ and $C$-band data in CASA with the CLEAN task using the Cotton-Schwab algorithm in the Multi Frequency Synthesis mode and Briggs weighting.  For the $C$-band observations we smoothed the upper 7.45~GHz baseband to match the resolution of the lower 5~GHz baseband using the CASA task IMSMOOTH.  We used the JMFIT task in AIPS to fit a single elliptical Gaussian component to the emission to determine its flux density, extent and position.  Observational parameters are summarized in Table~\ref{tbl-1} and radio images with contours are shown in Figure~\ref{fig:radio}.  The arcsecond ($\sim$50~pc-scale) resolution observations at $L$-band reveal slightly resolved emission (Table~\ref{tbl-1}).  The sub-arcsecond ($\sim$6~pc-scale) resolution observations at $C$-band reveal modestly resolved emission (Table~\ref{tbl-1}) consistent with the location of a central, hard X-ray point source \citep{binder11} and the optical position of the nucleus \citep{seth10}, as shown in Figure~\ref{fig:overlay}.  The 5~GHz emission is centered at $\alpha_{\mathrm{J2000}}$~=~01$^{\mathrm{h}}$09$^{\mathrm{m}}$27.00$^{\mathrm{s}}$ and $\delta_{\mathrm{J2000}}$~=~+35$\degr$43$\arcmin$04.91$\arcsec$ (with a 1-D positional uncertainty of 0.1$\arcsec$ dominated by calibration strategies).  From the integrated flux densities at 5 and 7.45~GHz, we obtain a $C$-band spectral index of $\alpha$~=~$-$0.88~$\pm$~0.30 (where $\mathrm{S}\varpropto \mathrm{\nu}^{\mathrm{\alpha}}$).  We also used our 1.5~GHz flux density to compute the three-point radio spectral index of $\alpha$~=~$-$1.15~$\pm$~0.05.  However, we caution that this provides only an estimate of the spectral index between 1.5 and 7.5~GHz since the 1.5~GHz data was an order of magnitude lower in resolution and not observed simultaneously with the 5 and 7.45~GHz data.







\section{Discussion}
\label{sec:results}
While the parsec-scale radio emission in the NGC~404 nucleus may originate from low-level accretion onto a central IMBH, this is not the only possibility.  The radio emission could also conceivably be produced by a single XRB, a young supernova remnant (SNR) or a recent bout of SF.  We utilize the ratio of the 5~GHz radio luminosity to the hard (2$-$10 keV) X-ray luminosity, $R_\mathrm{X}$ = $\mathrm{\nu}L_{\mathrm{\nu}}$(5 GHz)/$L_\mathrm{X}$ \citep{terashima03}, to help distinguish among the different possibilities for the origin of the radio and X-ray emission.  For NGC~404, we find $\log R_\mathrm{X}$~=~$-$2.5.  We discuss the possibility and limitation of this ratio being explained by an X-ray binary ($\S$3.1), star formation ($\S$3.2), a supernova remnant ($\S$3.3) and our preferred interpretation of an accreting IMBH ($\S$3.4).

\subsection{X-ray Binary}
Although the X-ray luminosity of the compact, central source in NGC~404 of $\sim$~1.2~$\times$~10$^{37}$~erg~s$^{-1}$ could be generated by a single XRB \citep{binder11}, such objects are known to be radio-faint.  To our knowledge, radio emission from an XRB beyond the Milky Way has yet to be detected with certainty \citep{muxlow10}.  Even a bright radio flare from a galactic XRB such as Cyg~X-3 (D~=~7~Kpc), which recently flared to $\sim$20~Jy at 15~GHz \citep{corbel12}, has a radio power of only $\sim$~1.2~$\times$~10$^{17}$ W~Hz$^{-1}$.  This is still nearly a factor of two less than the radio power of NGC~404 extrapolated to 15~GHz conservatively using our steeper, three-point radio spectral index of $\alpha$~=~$-1.15$.  Thus, the radio luminosity of NGC 404 is too high to have been generated by a typical XRB, let alone an XRB undergoing an extreme flaring event.  This characteristic radio faintness of XRBs also manifests itself in their radio/X-ray ratios, which tend to be much lower than the radio/X-ray ratio of NGC~404.  For instance, the radio/X-ray ratios of a sample of 7 galactic XRBs of various types have $\log R_\mathrm{X}$~$< -$5.3 \citep{merloni03, reines11}.  Therefore, an XRB origin cannot explain both the observed radio and X-ray emission in the NGC~404 nucleus. 

\subsection{Star Formation}
Evidence for recent nuclear SF in the center of NGC~404 includes the presence of the NSC; young, massive stars \citep{maoz05} and diffuse, soft X-ray emission \citep{binder11}.  Seth et al.\ (2010) estimated a current star formation rate (SFR) upper limit of 1.0~$\times$~10$^{-3}$~M$_{\sun}$~yr$^{-1}$ based on extinction-corrected nuclear H$\mathrm{\alpha}$ measurements from a 1$\arcsec$ (15~pc) slit on the 6.5m MMT.  We used the following relation to estimate the total (thermal + nonthermal) radio flux expected from the current SFR at 1.5 and 5~GHz \citep{murphy11}:

\begin{eqnarray}
\label{eq:murphy_total}
\left( \frac{\mathrm{SFR_{\mathrm{\nu}}}}{M_{\sun}\, \mathrm{yr^{-1}}} \right) = 10^{-27} \left[   2.18\left( \frac{T_{\mathrm{e}}} {10^4\, \mathrm{K}} \right)^{0.45} \left(  \frac{\mathrm{\nu}}{\mathrm{GHz}} \right)^{-0.1}  + 15.1\left( \frac{\mathrm{\nu}}{\mathrm{GHz}} \right)^{\mathrm{\alpha}_{\mathrm{NT}}}   \right]^{-1}  \nonumber \\ 
 \times\, \left(  \frac{L_{\mathrm{\nu}}}{\mathrm{erg\, s^{-1}\, Hz^{-1}}} \right).
\end{eqnarray}

\noindent We evaluated Equation \ref{eq:murphy_total} with $T_\mathrm{e}$~=~10$^4$~K and $\mathrm{\alpha_{NT}}$~=~$-$0.8 \citep{murphy11} and found upper limits on the 1.5 and 5~GHz contributions to SF of 1.3~$\times$~10$^{25}$~erg~s$^{-1}$~Hz$^{-1}$ (1.3~$\times$~10$^{18}$~W~Hz$^{-1}$) and 5.7~$\times$~10$^{24}$~erg~s$^{-1}$~Hz$^{-1}$ (5.7~$\times$~10$^{17}$~W~Hz$^{-1}$), respectively.  These luminosities correspond to flux densities of 1.1~mJy at 1.5~GHz and 0.5~mJy at 5~GHz.  These estimates indicate that a SF origin for the observed radio emission is indeed plausible, contrary to the conclusion of Binder et al.\ (2011).  However, we point out that the predicted upper limit on the 1.5~GHz flux due to SF is over a factor of 2.5 lower than our measured 1.5~GHz flux of 2.83~mJy.  The predicted upper limit on the 5~GHz flux due to SF is also lower than the 5~GHz flux density measured at parsec-scale resolution.  This suggests that an additional emission source, such as a weak LLAGN, would be needed to provide the missing flux density.

\subsection{Supernova Remnant}
A single, young supernova remnant (SNR) could also be responsible for the radio emission.  For instance, if the galactic SNR Cas~A were moved to the distance of NGC~404, it would have $S_{1.4\, \mathrm{GHz}}$~=~2.8~mJy and $S_{5\, \mathrm{GHz}}$~=~1.2~mJy \citep{baars77}, roughly similar to the observed radio flux densities in NGC~404.  The spatial extent of a Cas~A-like SNR in NGC~404 would also be similar to the observed 5~GHz radio size.  At an age of about 300~years, Cas~A has a diameter of $\sim$5~pc, similar to the observed $\sim$8~pc extent of the emission in NGC~404.  The well-measured radio spectral index of Cas~A of $\mathrm{\alpha}$~=~$-$0.77 is also similar to the $C$-band spectral index of $\mathrm{\alpha}$~=~$-$0.88 for NGC~404.  However, the radio/X-ray ratio of Cas~A is only $\log R_\mathrm{X}$~$\sim$~$-$1.7 \citep{reines11} since SNRs tend to be relatively weak in the hard X-ray regime.

An utraluminous SNR, such as J1228+441 in the nearby galaxy NGC~4449, is another potential culprit for the X-ray and radio emission.  J1228+441 is located at a distance of 3.8~Mpc \citep{annibali08} and is 60~$-$~200~years old \citep{lacey07}.  Like typical young SNRs, J1228+441 is compact with a diameter upper limit of 0.028$\arcsec$~=~0.5~pc.  Over three decades of radio observations, the flux densities and radio spectral indices of this young SNR have varied, with 5~GHz luminosities ranging from 2.2~$\times$~10$^{26}$~erg~s$^{-1}$~Hz$^{-1}$ to 6.9~$\times$~10$^{25}$~erg~s$^{-1}$~Hz~$^{-1}$ and a radio spectral index ranging from $-$0.6 to $-$1.0 from 1973 to 2002 \citep{lacey07}.  The ratio of the 5~GHz luminosity measured in 2002 to the 2$-$10~keV luminosity measured by $Chandra$ in 2001 of $\sim$1.7~$\times$~10$^{38}$~erg~s$^{-1}$ \citep{summers03} gives $\log R_\mathrm{X}$~$\sim$~$-$2.7 \citep{reines11}, which is very similar to our $\log R_\mathrm{X}$ measurement in NGC~404.  Although J1228+441 and NGC 404 share similar radio/X-ray ratios, we point out that many properties of the ultraluminous SNR in question are the result of the highly dense ISM in the host galaxy, NGC 4449.  Therefore, such an extreme type of SNR is not expected in a galaxy with no evidence of a dense ISM, such as NGC 404 \citep{kuno02}. 

Although an SNR origin for the radio emission remains a possibility, it is not the most likely scenario given the low SFR in the NGC~404 nucleus.  We used the following relation with the SFR upper limit of 1~$\times$~10$^{-3}$~M$_{\sun}$~yr$^{-1}$ \citep{seth10} to estimate the supernova rate, $\dot{N}_{SN}$ \citep{murphy11}:

\begin{equation}
\label{eq:snrate}
\left( \frac{\mathrm{SFR}}{M_{\sun}\, \mathrm{yr^{-1}}} \right) = 86.3 \left( \frac{\dot{N}_{\mathrm{SN}}}{\mathrm{yr}^{-1}} \right)
\end{equation}

\noindent and found $\dot{N}_{SN}$~$\approx$~1~$\times$~10$^{-5}$~yr$^{-1}$.  The Poisson distribution gives a probability of $\sim$1\% that at least one supernova has occurred in the nuclear region of NGC~404 in the last 1000~years.  Given the low probability for a recent supernova, we conclude that a single, young SNR is not the most plausible explanation for the observed radio emission.  Future Very Long Baseline Interferometric (VLBI) imaging with milliarcsecond-scale resolution would be able to more definitively rule out an SNR origin for the radio emission.

\subsection{Accreting Intermediate-Mass Black Hole}
An IMBH-driven LLAGN origin for the radio and X-ray emission is a tantalizing possibility.  The $C$-band spectral index of NGC~404 of $\alpha$~=~$-$0.88 is consistent with the wide range of radio spectral indices observed for LLAGNs \citep{hoandulvestad01}.  The $\log R_\mathrm{X}$ value of NGC~404 of $-$2.5 is formally radio-loud, where radio-loud is defined as $\log$~R$_{\mathrm{X}}$~$>$~$-$4.5 \citep{terashima03}.  This is consistent with the typically radio-loud nature of LINERs and LLAGNs in general \citep{terashima03, ho2008}.  The median $\log$~R$_{\mathrm{X}}$ for all LLAGN subclasses ranges from $-$2.8 to $-$3.8, and the median value for type 2 LINERs specifically is $-$2.9 \citep{ho2008}.  It has been shown that as radio-loudness increases in LLAGNs, the Eddington ratio decreases \citep{ho2002, terashima03, maccarone04}.  The Eddington ratio for the IMBH candidate in NGC~404, assuming a mass of 4.5~$\times$~10$^{5}$~M$_{\sun}$, is $L_{\mathrm{Bol}}$/$L_{\mathrm{Edd}}$~$\sim$~3~$\times$~10$^{-6}$ (where $L_{\mathrm{X}}$~=~$L_{\mathrm{Bol}}$/16, Ho 2008).  Thus, the IMBH in the center of NGC~404 is radio-loud and accreting well below its Eddington limit, similar to other weak LLAGNs.  A comparison of  the radio and X-ray properties of NGC~404 with other prominent IMBH candidates is provided in Table~\ref{tbl-2}.

If we assume that the NGC~404 nucleus indeed harbors a BH, we can obtain a BH mass estimate by utilizing the ``Fundamental Plane of  Black Hole Activity" (FPBHA) \citep{merloni03, falcke04, kording06}, which relates 5~GHz radio luminosity, 2$-$10~keV X-ray luminosity and BH mass for all BHs ranging from stellar-mass to SMBHs.  The FPBHA is defined by:
\begin{equation}
\label{eq1}
\log(L_{\mathrm{R}}) = \xi_{\mathrm{RX}} \log( L_{\mathrm{X}})+ \xi_{\mathrm{RM}} \log(M) + b_{\mathrm{R}}
\end{equation}
where the units of $L_{\mathrm{X}}$ are erg~s$^{-1}$, the units of $M$ are M$_{\sun}$ and $L_{\mathrm{R}}$~=~${\mathrm{\nu}}L_{\mathrm{\nu}}$~erg~s$^{-1}$.  

The values of the empirically-derived correlation coefficients depend on the sample selection \citep{kording06} and can provide insights into the underlying accretion physics.  The Merloni et al.\ (2003) correlation coefficients are $\xi_{\mathrm{RX}}$~= ~$0.60^{+0.11}_{-0.11}$, $\xi_{\mathrm{RM}}$~=~$0.78^{+0.11}_{-0.09}$ and $b_{\mathrm{R}}$~=~$7.33^{+4.05}_{-4.07}$.  These correlation coefficients, along with the measured radio and X-ray luminosities of the IMBH candidate in the center of NGC~404, give a BH mass estimate of log($M$)~$\sim$~6.4~$\pm$~1.1~M$_{\sun}$ (2.5$^{+29.1}_{-2.3}$~$\times$~10$^6$~M$_{\sun}$).  This BH mass estimate is consistent, within the wide range of uncertainties, with the dynamical BH mass estimate of $\sim$~4.5~$\times$~10$^{5}$~M$_{\sun}$ \citep{seth10} as well as the BH mass predicted by the $M_{\mathrm{BH}}-\sigma_{*}$ relation of a few~$\times$~10$^5$~M$_{\sun}$ for $\sigma_{*}~\sim$~40~km~s$^{-1}$ \citep{gultekin09, seth10}.

\section{Summary}
\label{sec:summary}
We observed the candidate IMBH in the NGC~404 nucleus with the EVLA in the B-configuration at $L$-band (1-2~GHz) and the A-configuration at $C$-Band (4-8~GHz) and detected, for the first time, radio emission at a frequency higher than 1.4~GHz.  The $L$-band emission is slightly resolved on arcsecond ($\sim$50~pc) scales at 1.5~GHz and has a radio power of 3.3~$\pm$~0.2~$\times$~10$^{18}$~W~Hz$^{-1}$.  The 5~GHz emission is modestly resolved on subarcsecond ($\sim$6~pc) scales with a radio power of 7.6~$\pm$~0.7~$\times$~10$^{17}$~W~Hz$^{-1}$ and a $C$-band spectral index from 5 to 7.45~GHz of $\alpha$~=~$-$0.88~$\pm$~0.30.  NGC~404 is only the third IMBH candidate with detected subarcsecond-scale radio continuum emission.  The radio luminosity, spectral index and radio/X-ray ratio in the NGC~404 nucleus are most consistent with an IMBH-powered LLAGN.  The location of the radio emission is consistent with the position of a hard X-ray point source \citep{binder11} and the optical center of the galaxy.  An XRB origin is unlikely given NGC~404's high 5~GHz luminosity and radio/X-ray ratio (log$R_{\mathrm{X}}$ = $-$2.5).  A single, young SNR comparable to Cas~A in the NGC~404 nucleus could produce emission similar to the observed 5~GHz emission, but it could not explain the observed hard X-ray emission.  An ultraluminous SNR similar to J1228+441 could conceivably produce the observed radio and X-ray emission, however, given the low supernova rate and lack of evidence for a dense ISM, the probability of this scenario is low.  A SF origin for the emission remains a plausible scenario, although the upper limit on the predicted 5~GHz flux density due to SF is less than that observed.  The upper limit on the SF contribution to the 1.5~GHz emission was deficient by over a factor of 2.5 compared to our measured 1.5~GHz flux, further suggesting that SF alone may not be able to produce all of the observed emission.  In summary, alternative explanations for the observed radio and X-ray emission exist but are either inconsistent with the observed X-ray and/or radio luminosities or are statistically unlikely.  Furthermore, our FPBHA \citep{merloni03} BH mass estimate is compatible with the gas dynamical mass estimate \citep{seth10} and $M_{\mathrm{BH}}-\sigma_{*}$ relation estimates \citep{gultekin09, seth10}.  Thus, to date, the most likely scenario for the compact 5~GHz radio and hard X-ray emission in the nucleus of NGC~404 is an IMBH accreting at low levels.
 
A future EVLA study with the dedicated goal of mapping the radio spectral index at a variety of frequencies would allow us to better model the underlying emission as well as monitor the source for radio variability.  Milliarcsecond-scale resolution VLBI imaging of the NGC~404 nucleus would provide a robust measurement of the brightness temperature capable of definitively ruling out a compact starburst, and may also reveal morphological features such as jets characteristic of some LLAGNs \citep{nagar05}.  With ongoing sensitivity improvements at the Very Long Baseline Array, such observations of NGC~404 may be possible in the near future.  Continued studies of the nuclei of nearby, low-mass galaxies such as NGC~404 will ultimately help improve our understanding of the interplay between BHs and galaxy evolution and may even provide deeper insight into the long-debated issue of the relationship between AGN feedback and nuclear SF. 

\acknowledgments
We thank the anonymous referee for helpful suggestions that have improved the clarity and strength of this work.  We also thank Anil Seth for helpful comments and for providing us with coordinate frame-corrected Hubble Legacy Archive images.  The National Radio Astronomy Observatory is a facility of the National Science Foundation operated under cooperative agreement by Associated Universities, Inc.  This research was partially funded by the NRAO Graduate Summer Student Research Assistantship Program.\

{\it Facilities:} \facility{NRAO}.

\clearpage

\clearpage
\begin{deluxetable}{ccccccrccc}
\tabletypesize{\scriptsize}
\tablecolumns{10}
\tablewidth{0pt}
\rotate
\tablecaption{Summary of EVLA Observations of NGC 404 \label{tbl-1}}
\tablehead{
\colhead{} & \colhead{} & \multicolumn{2}{c}{Beam Parameters} & \colhead{}  & \colhead{} & \colhead{}& \colhead{Source Parameters} & \colhead{} & \colhead{}  \\
\cline{3-4} \cline{6-10} \\
\colhead{Frequency} & \colhead{rms} & \colhead{$\theta_{M} \times \theta_{m}$} & \colhead{P.A.} & \colhead{} & \colhead{$\theta_{M} \times \theta_{m}$} & \colhead{P.A.} & \colhead{$M \times m$} & \colhead{Total Flux Density} & \colhead{log($P_{\mathrm{rad}}$)}\\

 \colhead{(GHz)}  &  \colhead{($\mu$Jy beam$^{-1}$)}  &  \colhead{($\arcsec$)}  &  \colhead{(deg)} & \colhead{} & \colhead{($\arcsec$)} & \colhead{(deg)} & \colhead{(pc)} &  \colhead{(mJy)}  &  \colhead{(W Hz$^{-1}$)} \\
 
   \colhead{(1)} &  \colhead{(2)}  &  \colhead{(3)}  &  \colhead{(4)} & \colhead{}  &  \colhead{(5)} & \colhead{(6)} &  \colhead{(7)}  & \colhead{(8)} & \colhead{(9)}
}
\startdata
 7.45 & 12 & 0.39 $\times$ 0.32 & 103 & & 0.54 $\pm$ 0.05 $\times$ 0.36 $\pm$ 0.06 & 19.7   $\pm$ 10.0  &   8.1 $\pm$ 0.8 $\times$ 5.4   $\pm$ 0.9 & 0.47 $\pm$ 0.04 & 17.73 \\
      5  & 16 & 0.39 $\times$ 0.32 & 103 & & 0.56 $\pm$ 0.05 $\times$ 0.35 $\pm$ 0.05 & 178.2 $\pm$ 13.4 &   8.4 $\pm$ 0.8 $\times$ 5.3    $\pm$ 0.8 & 0.66 $\pm$ 0.06 & 17.88 \\   
    1.5 & 55 & 3.42 $\times$ 2.79 & 122 & & 2.32 $\pm$ 0.22 $\times$ 1.33 $\pm$ 0.26 & 101.5 $\pm$ 10.7 & 34.9 $\pm$ 3.3 $\times$ 20.0 $\pm$ 3.9 & 2.83 $\pm$ 0.14 & 18.51\\
\enddata
\tablecomments{Col. (1): Center frequency.  Col. (2): Average rms noise in image.  Col. (3): Clean beam major $\times$ minor axis.  Col. (4): Clean beam position angle. Col. (5): Angular dimensions (major $\times$ minor axis) of the deconvolved emission and error bars from JMFIT in AIPS.  Col. (6): Position Angle of the deconvolved emission from JMFIT in AIPS.  Col. (7): Linear dimensions (major $\times$ minor axis) of the deconvolved emission assuming a distance of 3.1 Mpc.  Col. (8): Total integrated flux density and error.  The error shown is the sum of the error reported by JMFIT in AIPS and the 3\% calibration error, added in quadrature.  Col. (9): Log of the radio power assuming a distance of 3.1 Mpc.}
\end{deluxetable}

\clearpage
\begin{deluxetable}{ccccccccccccccc}
\tabletypesize{\scriptsize}
\rotate
\tablecaption{IMBH Candidates with High-Resolution Radio and X-ray Data \label{tbl-2}}
\tablewidth{0pt}
\tablehead{
\colhead{Name} & \colhead{Morph.} & \colhead{D} & \colhead{Ref.}  & \colhead{log($M_{\mathrm{BH}}$)} & \colhead{Method} & \colhead{Ref.} &
\colhead{log($L_{\mathrm{R}}$)} & \colhead{$\mathrm{\alpha}$} & \colhead{Ref.} & \colhead{log($L_{\mathrm{X}}$)} & \colhead{Ref.} & \colhead{log($R_{\mathrm{X}}$)} & \colhead{$L_{\mathrm{Bol}}$/$L_{\mathrm{Edd}}$} & \colhead{Ref.}\\
 \colhead{} &  \colhead{}  &  \colhead{(Mpc)} & \colhead{} &  \colhead{(M$_{\sun}$)} & \colhead{} & \colhead{} &
  \colhead{(erg s$^{-1}$)} &  \colhead{} &  \colhead{}  &  \colhead{(erg s$^{-1}$)}  &  \colhead{} &  \colhead{} &  \colhead{} &  \colhead{}\\
   \colhead{(1)} &  \colhead{(2)}  &  \colhead{(3)}  &  \colhead{(4)}  &  \colhead{(5)} &
  \colhead{(6)} &  \colhead{(7)}  & \colhead{(8)} & \colhead{(9)} & \colhead{(10)} & \colhead{(11)} & \colhead{(12)} & \colhead{(13)} & \colhead{(14)} & \colhead{(15)}
}
\startdata
NGC 404                                     &  dS0          & 3.1      &  1, 2   & 5.65 $\pm$ 0.25                                 & I   & 7    & 34.58 & -0.88 & 10 & 37.10 & 13 & -2.5 & -5.5 & 10\\
NGC 4395                                   & Sdm          & 4.3      &  3       & 5.56 $\pm$ 0.12                                 & II  & 8    & 34.95 & -0.60 & 11 & 39.91 & 14 & -5.0 & -2.9 & 8\\
GH 10                                           &  \nodata   & 363.0 &   4, 5   & 5.93 $\pm$ 0.50\tablenotemark{b} & III & 4    & 38.76 & -0.69 & 12 & 42.28 & 5  & -3.5 & 0.3 & 4\\
Henize 2-10\tablenotemark{a}& BCD         & 9.0      &  6        & 6.3 $\pm$ 1.1                                      & IV& 9    & 35.87 & -0.39 & 9 &  39.43 &  9   & -3.6 & -4.0 & 9\\
\enddata
\tablenotetext{a}{The central black hole mass estimate for Henize 2-10 is above 10$^6$ M$_{\sun}$ and it is therefore not, strictly speaking, an IMBH candidate.  However, the mass estimate is quite uncertain and the lower limit of the estimate lies within the IMBH range.}
\tablenotetext{b}{No uncertainty was given for the estimate of $M_{\mathrm{BH}}$ in GH~10 in Greene \& Ho (2004).  However, the authors stated that their estimated black hole virial masses have a scatter of $\sim$0.5~dex around the $M_{\mathrm{BH}} - \sigma_{*}$ relation, and we have adopted this value as a rough estimate of the uncertainty.}
\tablecomments{Col. (2): Galaxy morphology.  Col. (3): Distance.  Col. (4): Distance reference. Col. (5): Log of the black hole mass.  Col. (6): Method of black hole mass estimation.  I: dynamical measurement, II: reverberation mapping, III: H$\mathrm{\alpha}$ line width-luminosity mass scaling relation and IV: Fundamental Plane of Black Hole Activity.  Col. (7): Black hole mass reference.  Col. (8): Log of the 5~GHz radio luminosity (where $L_{\mathrm{R}}$ = $\mathrm{\nu}L_{\mathrm{\nu}}$(5~GHz)).  Col. (9): Radio spectral index, defined by $\mathrm{S}\varpropto \mathrm{\nu}^{\mathrm{\alpha}}$.  For NGC~404 the radio spectral index is between 5 and 7.45~GHz, for NGC~4395 and GH~10 it's between 1.4 and 5~GHz and for Henize~2-10 it's between 5 and 8.5~GHz.  Col. (10): Radio data reference.  Col. (11): Log of the 2$-$10 keV X-ray luminosity.  Col. (12): X-ray luminosity reference.  Col. (13): Log of the radio/X-ray ratio, defined as $R_\mathrm{X}$ = $\mathrm{\nu}L_{\mathrm{\nu}}$(5 GHz)/$L_\mathrm{X}$.  Col. (14): Log of the Eddington ratio.  Col. (15): Eddington ratio reference.
}  
\tablerefs{
(1) Karachentsev et al.\ 2002; 
(2) Dalcanton et al.\ 2009; 
(3) Thim et al.\ 2004; 
(4) Greene \& Ho 2004; 
(5) Greene \& Ho 2007a
(6) Johnson et al.\ 2000; 
(7) Seth et al.\ 2010; 
(8) Peterson et al.\ 2005;
(9) Reines et al.\ 2011;
(10) This work;
(11) Ho et al.\ 2001;
(12) Wrobel et al.\ 2008;
(13) Binder et al.\ 2011;
(14) Moran et al.\ 2005
}
\end{deluxetable}


\begin{thebibliography}{}
\bibitem[Annibali et al.(2008)]{annibali08} Annibali, F., Aloisi, A., Mack, J., Tossi, M. van der Marel, R. P. Angeretti, L., Leitherer, C., Sirianni, M. 2008, \apj, 135, 1900 
\bibitem[Baars et al.(1977)]{baars77} Baars, J. W. M., Genzel, R., Pauliny-Toth, I. I. K, Witzel, A. 1977, A\&A, 61, 99 
\bibitem[Baldry et al.(2004)]{baldry04} Baldry, I. K., Glazebrook, K., Brinkmann, J.,  Ivezi\'{c}, \v{Z}., Lupton, R. H., Nichol, R. C., Szalay, A. S. 2004, \apj, 600, 681 
\bibitem[Barth et al.(2004)]{barth04} Barth, A. J., Ho, Luis, C., Rutledge, R. E. \& Sargent, W. L. W. 2004, \apj, 607, 90 
\bibitem[Barth et al.(2008)]{barth08} Barth, Aaron J., Greene, Jenny E., Ho, Luis C. 2008, \apj, 136, 1179 
\bibitem[Barth et al.(2009)]{barth09} Barth, Aaron J., Strigari, Louis E., Bentz, Misty C., Greene, Jenny E., Ho, Luis C. 2009, \apj, 690, 1031 
\bibitem[Binder et al.(2011)]{binder11} Binder, B., Williams, B. F., Eracleus, M., Seth, A. C., Dalcanton, J. J., Skillman, E. D., Weisz, D. R., Anderson, S. F., Gaetz, T. J., \& Plucinsky, P. P. 2011, \apj, 737, 77 
\bibitem[Bouchard et al.(2010)]{bouchard10} Bouchard, A., Prugniel, P., Koleva, M. \& Sharina, M. 2010, A\&A, 513, 54 
\bibitem[Corbel et al.(2012)]{corbel12} Corbel, S., Dubus, G., Tomsick, J. A., Szostek, A., Corbet, R. H. D., Miller-Jones, J. C. A., Richards, J. L., Pooley, G., Trushkin, S., Dubois, R., Hill, A. B., Kerr, M., Max-Moerbeck, W., Readhead, A. C. S., Bodaghee, A., Tudose, V., Parent, D., Wilms, J. and Pottschmidt, K. 2012, \mnras, accepted 
\bibitem[Dalcanton et al.(2009)]{dalcanton04} Dalcanton, J. J., Williams, B. F., Seth, A. C., Dolphin, A., Holtzman, J., Rosema, K., Skillman, E. D., Cole, A., Girardi, L., Gogarten, S. M., Karachentsev, I. D., Olsen, K., Weisz, D., Christensen, C., Freeman, K., Gilbert, K., Gallart, C., Harris, J., Hodge, P., de Jong, R. S., Karachentseva, V., Mateo, M., Stetson, P. B., Tavarez, M., Zaritsky, D., Governato, F., Quinn, T. 2009, \apjs, 183, 67 
\bibitem[del R\'{\i}o et al.(2004)]{delrio04} del R$\acute{\i}$o, M. S., Brinks, E., \& Cepa, J., 2004, \apj, 128, 89 
\bibitem[Desroches et al.(2009)]{desroches09} Desroches, Louis-Benoit, Greene, Jenny E., Ho, Luis C. 2009, \apj, 698, 1515 
\bibitem[Falcke(2004)]{falcke04} Falcke, H., Kording, E. \& Markoff, S. 2004, A\&A, 414, 895  
\bibitem[Farrell et al.(2012)]{farrell12} Farrell, S. A, Servillat, M., Pforr, J., Maccarone, T. J., Knigge, C., Godet, O., Maraston, C., Webb, N. A., Barret, D., Gosling, A. J., Belmont, R., Wiersema, K. 2012, \apj, 747, 13 
\bibitem[Ferrarese et al.(2006)]{ferrarese06} Ferrarese, L., C\^{o}t\'{e}, P., Dalla Bont\`{a}, E., Peng, E. W., Merritt, D., Jord\'{a}n, A., Blakeslee, J. P., Ha\c{s}egan, M., Mei, S., Piatek, S., Tonry, J. L, West, M. J. 2006, \apj, 644, 21 
\bibitem[Filippenko \& Ho(2003)]{filippenko03} Filippenko, Alexei V. \& Ho, Luis C. 2003, \apj, 588, 13 
\bibitem[Gebhardt et al.(2002)]{gebhardt02} Gebhardt, Karl, Rich, R. M, Ho, Luis C. 2002, \apj, 578, 41 
\bibitem[Gebhardt et al.(2005)]{gebhardt05} Gebhardt, Karl, Rich, R. M, Ho, Luis C. 2005, \apj, 634, 1093 
\bibitem[Graham \& Spitler(2009)]{graham09} Graham, A. W. \& Spitler, L. R. 2009, \apj, 397, 2148 
\bibitem[Greene et al.(2006)]{greene06} Greene, J. E., Ho, L. C., \& Ulvestad, J. S. 2006, \apj, 636, 56 
\bibitem[Greene \& Ho(2004)]{greene04} Greene, J. E., Ho, L. C. 2004, \apj, 610, 722 
\bibitem[Greene \& Ho(2007a)]{greene07a} Greene, J. E., Ho, L. C. 2007a, \apj, 656, 84 
\bibitem[Greene \& Ho(2007b)]{greene07b} Greene, J. E., Ho, L. C. 2007b, \apj, 670, 92 
\bibitem[Gultekin et al.(2009)]{gultekin09} Gultekin, K., Cackett, E. M., Miller, J. M., Di Matteo, T., Markoff, S., \& Richstone, D. O. 2009, \apj, 706, 404 
\bibitem[Ho(2002)]{ho2002} Ho, L. C. 2002, \apj, 564, 120 
\bibitem[Ho(2008)]{ho2008} Ho, L. C. 2008, \araa, 46, 475 
\bibitem[Ho et al.(1997)]{ho97} Ho, L. C., Filippenko, A. V., Sargent, W. L. W 1997, \apjs, 112, 315 
\bibitem[Ho \& Ulvestad(2001)]{hoandulvestad01} Ho, L. C. \& Ulvestad, J. S. 2001, \apjs, 133, 77 
\bibitem[Hughes(2009)]{hughes09} Hughes, S. A. 2009, \araa, 47, 107 
\bibitem[Ibata et al.(2009)]{ibata09} Ibata, R., Bellazzini, M., Chapman, S. C., Dalessandro, E., Ferraro, F., Irwin, M., Lanzoni, B., Lewis, G. F., Mackey, A. D., Miocchi, P., Varghese, A. 2009, \apj, 699, 169 
\bibitem[Johnson et al.(2000)]{johnson00}  Johnson, Kelsey E., Leitherer, Claus, Vacca, William D., Conti, Peter S. 2000, AJ, 120, 1273 
\bibitem[Karachentsev et al.(2002)]{karachentsev02} Karachentsev, I. D., Sharina, M. E., Makarov, D. I., Dolphin, A. E., Grebel, E. K., Geisler, D., Guhathakurta, P., Hodge, P. W., Karachentseva, V. E., Sarajedini, A., Seitzer, P. 2002, A\&A, 389, 812 
\bibitem[Kong et al.(2010)]{kong10} Kong, A. K. H., Heinke, C. O., Di Stefano, R., Cohn, H. N., Lugger, P. M., Barmby, P., Lewin, W. H. G., Primini, F. A. 2010, \mnras, 407, 84 
\bibitem[Kording et al.(2006)]{kording06} Kording, E., Falcke, H., Corbel, S. 2006, A\&A, 456, 439 
\bibitem[Kormendy \& Richstone(1995)]{kor_rich95} Kormendy, J. \& Richstone, R. 1995, \araa, 33, 581 
\bibitem[Kuno et al.(2002)]{kuno02}  Kuno, Nario, Nakai, Naomasa, Sorai, Kazuo, Nishiyama, Kohta, Vila-Vila\'{o}, Baltasar 2002, PASJ, 54, 555 
\bibitem[Lacey et al.(2007)]{lacey07} Lacey, C. K., Goss, W. M., Mizouni, L. K. 2007, \apj, 133, 2156 
\bibitem[Maccarone(2004)]{maccarone04} Maccarone, T. 2004, \mnras, 351, 1049 
\bibitem[Maoz et al.(1998)]{maoz98} Maoz, D., Koratkar, A. P., Shields, J. C., Ho, L. C., Filippenko, A. V., \& Sternberg, A. 1998, \aj, 116, 55 
\bibitem[Maoz et al.(2005)]{maoz05} Maoz, D., Nagar, N. M., Falcke, H. \& Wilson, A. S. 2005, AJ, 625, 699  
\bibitem[McAlpine et al.(2011)]{mcalpine11} McAlpine, W., Satyapal, S., Gliozzi, M., Cheung, C. C., Sambruna, R. M. \& Eracleous, Michael 2011, \apj, 728, 25 
\bibitem[McKernan et al.(2011)]{mckernan11} McKernan, B., Ford, K. E. S., Yaqoob, T., Winter, L. M. 2011, \mnras, 413, 24 
\bibitem[Merloni et al.(2003)]{merloni03} Merloni, A., Heinz, S. \& Di Matteo, T. 2003, \mnras, 345, 1057 
\bibitem[Moran et al.(2005)]{moran05}  Moran, Edward C., Eracleous, Michael, Leighly, Karen M., Chartas, George, Filippenko, Alexei V., Ho, Luis C., Blanco, Philip R. 2005, AJ, 129, 2108 
\bibitem[Murphy et al.(2011)]{murphy11} Murphy, E. J., Condon, J. J., Schinnerer, E., Kennicutt, R. C., Calzetti, D., Armus, L., Helou, G., Turner, J. L., Aniano, G. \& Beir\~{a}o, P. 2011, \apj, 737, 67 
\bibitem[Muxlow et al.(2010)]{muxlow10} Muxlow, T. W. B., Beswick, R. J., Garrington, S. T., Pedlar, A., Fenech, D. M., Argo, M. K., van Eymeren, J., Ward, M., Zezas, A., Brunthaler, A. 2010, \mnras, 404, 109
\bibitem[Nagar et al.(2005)]{nagar05} Nagar, N. M., Falcke, H., Wilson, A. S. 2005, A\&A, 435, 521 
\bibitem[Noyola et al.(2010)]{noyola10} Noyola, Eva, Gebhardt, Karl, Kissler-Patig, Markus, L$\mathrm{\ddot{u}}$tzgendorf, Nora, Jalali, Behrang, de Zeeuw, P. Tim, Baumagardt, Holger 2010, \apj, 719, 60 
\bibitem[Perley(2011)]{perley11} Perley, R. A., Chandler, C. J., Butler, B. J., Wrobel, J. M. 2011, \apj, 739, 1 
\bibitem[Peterson et al.(2005)]{peterson05} Peterson, Bradley M., Bentz, Misty C., Desroches, Louis-Benoit, Filippenko, Alexei V., Ho, Luis C., Kaspi, Shai, Laor, Ari, Maoz, Dan, Moran, Edward C., Pogge, Richard W. \& Quillen, Alice C. 2005, \apj, 632, 799. 
\bibitem[Ravindranath et al.(2001)]{ravindranath01} Ravindranath, S., Ho, L. C., Peng, C. Y., Filippenko, A. V. \& Sargent, W. L. W. 2001, \apj, 122, 653  
\bibitem[Reines et al.(2011)]{reines11} Reines, A. E., Sivakoff, G. R., Johnson, K. E. \& Brogan, C. L. 2011, Nature, 470, 66 
\bibitem[Satyapal et al.(2004)]{satyapal04} Satyapal, S., Sambruna, R. M. \& Dudik, R. P. 2004, A\&A, 414, 825 
\bibitem[Satyapal et al.(2009)]{satyapal09} Satyapal, S., B$\mathrm{\ddot{o}}$ker, T., McAlpine, W., Gliozzi, M., Abel, N. P., Heckman, T. 2009, \apj, 704, 439 
\bibitem[Seth et al.(2008)]{seth08} Seth, A., Ag\"{u}eros, M., Lee, D. \& Basu-Zych, A. 2008, \apj, 678, 116 
\bibitem[Seth et al.(2010)]{seth10} Seth, A. C., Cappellari, M., Neumayer, N., Caldwell, N., Bastian, N., Olsen, K., Blum, R. D., Debattista, V. P., McDermid, R., Puzia, T., Stephens, A. 2010, \apj, 714, 713 
\bibitem[Shields et al.(2008)]{shields08} Shields, Joseph C., Walcher, Jakob, B$\mathrm{\ddot{o}}$ker, Torsten, Ho, Luis C., Rix, Hans-Walter \& van der Marel, Roeland P. 2008, \apj, 682, 104 
\bibitem[Summers et al.(2003)]{summers03} Summers, L. K., Stevens, I. R., Strickland, D. K., Heckman, T. M. 2003, MNRAS, 342, 690
\bibitem[Terashima \& Wilson(2003)]{terashima03} Terashima, Y., Wilson, A. S. 2003, \apj, 583, 145 
\bibitem[Thilker et al.(2010)]{thilker10} Thilker, D. A., Bianchi, L., Schiminovich, D., Gil de Paz, A., Seibert, M., Madore, B. F., Wyder, T., Rich, R. M., Yi, S., Barlow, T., Conrow, T., Forster, K., Friedman, P., Martin, C., Morrissey, P., Neef, S. \& Small, T. 2010, \apjl, 714, 171 
\bibitem[Thim et al.(2004)]{thim04} Thim, F., Hoessel, J. G., Saha, A., Claver, J., Dolphin, A., Tammann, G. A. 2004, AJ, 127, 2322 
\bibitem[Thornton et al.(2008)]{thornton08} Thornton, C. E., Barth, A. J., Ho, L. C., Rutledge, R. E., Greene, J. E. 2008, \apj, 686, 892 
\bibitem[Tremaine et al.(2002)]{tremaine02} Tremaine, Scott, Gebhardt, Karl, Bender, Ralf, Bower, Gary, Dressler, Alan, Faber, S. M., Filippenko, Alexei V., Green, Richard, Grillmair, Carl, Ho, Luis C., Kormendy, John, Lauer, Tod R., Magorrian, John, Pinkney, Jason, Richstone, Douglas 2002, \apj, 574, 740 
\bibitem[Ulvestad et al.(2007)]{ulvestad07} Ulvestad, James S., Greene, Jenny E. \& Ho, Luis C. 2007, \apj, 661, 151 
\bibitem[van der Marel(2004)]{vandermarel04} van der Marel, R. P. 2004, in Coevolution of Black Holes and Galaxies, ed. L.C. Ho (Cambridge: Cambridge Univ. Press), 37 
\bibitem[van der Marel \& Anderson(2010)]{vandermarel10} van der Marel, Roeland, P. \& Anderson, Jay 2010, \apj, 710, 1063 
\bibitem[van Wassenhove et al.(2010)]{vanwass10} van Wassenhove, S., Volonteri, M., Walker, M. G., Gair, J. R. 2010, \mnras, 408, 1139 
\bibitem[Volonteri(2010)]{volonteri10} Volonteri, M. 2010, \araa, 18, 279 
\bibitem[Wrobel et al.(2001)]{wrobel01} Wrobel, J. M., Fassnacht, C. D. \& Ho, L. C. 2001, \apj, 553, 23 
\bibitem[Wrobel \& Ho(2006)]{wrobelandho06} Wrobel, J. M. \& Ho, L. C. 2006, \apj, 646, 95 
\bibitem[Wrobel et al.(2008)]{wrobel08} Wrobel, J. M., Greene, J. E., Ho, L. C. \& Ulvestad, J. S. 2008, \apj, 686, 838 
\bibitem[Wrobel et al.(2011)]{wrobel11} Wrobel, J. M., Greene, J. E., Ho, L. C. 2011, AJ, 142, 113 

\end{thebibliography}
\end{document}